\documentclass[12pt,thmsa]{article}

\begin{document}

\noindent \textbf{ABOUT\ THE\ SIMULTANEOUS\ CO-EXISTENCE\ OF\ INSTANTANEOUS\
AND\ RETARDED\ INTERACTIONS\ IN\ CLASSICAL\ ELECTRODYNAMICS \bigskip
\medskip }

LjUBA \v {S}KOVRLJ$^{*}$ and TOMISLAV IVEZI\'{C}$^{\dagger ,\ddagger }$

$^{*}$Faculty of Textile Technology, University of Zagreb, Pierottieva 6,
10000

Zagreb, Croatia

$^{\dagger }$Ru%
\mbox
 {\it{d}\hspace{-.15em}\rule[1.25ex]{.2em}{.04ex}\hspace{-.05em}}er Bo\v
{s}kovi\'{c} Institute, P.O.B. 180, 10002 Zagreb, Croatia

$^{\ddagger }$E-mail: ivezic@rudjer.irb.hr\bigskip \medskip

\noindent In this paper it is proved that, contrary to the results found by
A.E. Chubykalo and S.J. Vlaev (\textit{Int. J. Mod. Phys.} \textbf{14}, 3789
(1999)), the retarded electric and magnetic fields for an uniformly
accelerated charge exactly satisfy Maxwell equations (ME). Furthermore it is
shown that ME are correctly written in the usual form with the partial
derivatives and thus not, as proposed by Chubykalo and Vlaev, with the total
derivatives. \bigskip

\noindent \textbf{1.} \textbf{Introduction\medskip }

\noindent Recently Chubykalo and Vlaev$^{1}$ argued that the electromagnetic
fields (obtained from the Li\'{e}nard-Wiechert potentials (LW)) of a charge
moving with a constant acceleration along an axis do not satisfy Maxwell
equations (ME) if one considers exclusively a retarded interaction.
According to their interpretation the retarded interaction is mathematically
described by the implicit functional dependence of the electric $\mathbf{E}$
and magnetic $\mathbf{B}$ fields on the observation time $t.$ Further they
stated that ME will be satisfied ''if and only if'' one takes into account
an \emph{explicit} functional dependence of $\mathbf{E}$ and $\mathbf{B}$ on
the observation time $t$ \emph{together} with the mentioned implicit
dependence. This explicit dependence of fields on $t$ then mathematically
describes an instantaneous interaction. The same assertions were also
reported in series of papers, e.g., Refs. 2-4.

In the paper$^{5}$ we have argued that the ''electrodynamics dualism
concept''$^{4}$ with the simultaneous coexistence of instantaneous and short
range interaction is not correct. In this paper we show that the opinion
presented in$^{1-4}$ about the incorrectnes of the retarded solutions
emerges, in fact, from an incorrect mathematical procedure in treating the
partial derivatives of continuos functions of several variables. Their$%
^{1-4} $ physical interpretation with the ''electrodynamics dualism
concept'' is then a consequence of such an incorrect mathematical treatment.
Further it will be proved that, contrary to the assertions from,$^{1-4}$ the
ME
\begin{eqnarray}
\nabla \mathbf{E} &=&\rho /\epsilon _{0}  \nonumber \\
\nabla \mathbf{B} &=&0  \nonumber \\
\nabla \times \mathbf{B} &=&\mu _{0}\mathbf{j+}\mu _{0}\varepsilon
_{0}\partial \mathbf{E}/\partial t  \label{maxw} \\
\nabla \times \mathbf{E} &=&-\partial \mathbf{B}/\partial t  \nonumber
\end{eqnarray}
are correctly written in the usual form, Eqs. (\ref{maxw}), i.e., with the
partial derivatives and not with the total derivatives, and that the
retarded fields of a charge $q$ moving with a constant acceleration along
the $X$ axis, Eqs. (22)-(27) from Ref. 1, exactly satisfy ME (\ref{maxw}).
\bigskip

\noindent \textbf{2. Derivation of the Fields }$\mathbf{E}$\textbf{\ and }$%
\mathbf{B}$\textbf{\ Taking into Account the Retarded Interaction
Only\medskip }

\noindent First it will be discussed the calculation of the LW fields $%
\mathbf{E}$ and $\mathbf{B}$ from a charge $q$ for which the retarded
position $x_{0i}(t_{0})$ as a function of the retarded time $t_{0}$ is
assumed given. These fields are
\begin{eqnarray}
\mathbf{E} &=&Kq\left\{ (\mathbf{R}-\mathbf{V}R/c)(1-V^{2}/c^{2})/s^{3}%
\right\} _{t_{0}}+  \nonumber \\
&+&Kq\left\{ \left[ \mathbf{R\times }\left( (\mathbf{R}-R\mathbf{V}/c)\times
\mathbf{\dot{V}}\right) \right] /c^{2}s^{3}\right\} _{t_{0}},  \label{efiel}
\end{eqnarray}
\begin{equation}
\mathbf{B=}\left\{ \left[ \mathbf{R\times E}\right] /Rc\right\} _{t_{0}},
\label{bfiel}
\end{equation}
where $K=1/4\pi \varepsilon _{0}$ and $s=R-\mathbf{VR}/c.$ (The notation is
similar to that one in Ref. 1; $\left\{ ...\right\} _{t_{0}}$ means that all
functions of $x,y,z,t$ in the braces $\left\{ {}\right\} $ are taken at the
moment of time $t_{0},$ that is determined from the retardation condition (%
\ref{retco}), see below.) $\mathbf{E}$ and $\mathbf{B}$ are determined in
the usual way
\begin{equation}
\mathbf{E}=-\nabla \varphi -\partial \mathbf{A/}\partial t,\quad \mathbf{B=}%
\nabla \times \mathbf{A}  \label{eandb}
\end{equation}
from the LW potentials
\begin{equation}
\varphi \left( \mathbf{r},t\right) =\left\{ Kq/s\right\} _{t_{0}},\quad
\mathbf{A}\left( \mathbf{r},t\right) =\left\{ Kq\mathbf{V}/s\right\}
_{t_{0}}.  \label{poten}
\end{equation}
We note that the field and source point variables are connected by the
retardation condition
\begin{equation}
R\left( x_{i},x_{0i}\right) =\left[ \sum_{i}\left( x_{i}-x_{0i}\right)
^{2}\right] ^{1/2}=c\left( t-t_{0}\right) .  \label{retco}
\end{equation}
This equation is Eq. (8) from Ref. 1, see also Ref. 7 Sec. 20-1, or Ref. 6
Secs. 62 and 63. Since $x_{0i}$ is asumed given as a function of $t_{0}$ we
can write
\begin{equation}
R\left( x_{i},x_{0i}(t_{0})\right) =f\left( x,y,z,t_{0}\right) =c\left(
t-t_{0}\right) ,  \label{retco2}
\end{equation}
which is a functional relation between $x_{i}$, $t$ and $t_{0},$ see$^{7}$
Eq. (20-4). The LW potentials are expressed in terms of $R$ (\ref{retco}),
which means that $\varphi \left( \mathbf{r},t\right) $ and $\mathbf{A}\left(
\mathbf{r},t\right) $ are not only implicit functions of $x,y,z$ through $%
t_{0}$ but \emph{the explicit functions as well}. Although Landau and
Lifshitz$^{6}$ p. 161 stated that the LW potentials (\ref{poten}) are only
implicit functions of $x,y,z$ through $t_{0}$ they really calculated the
fields $\mathbf{E}$ and $\mathbf{B}$ taking into account that R, (\ref{retco}%
) or (\ref{retco2}), is also the explicit function of the field coordinates $%
x,y,z$ in the same way as in Ref. 7. This can be clearly seen from the
calculation of $\nabla t_{0}$ when $\nabla R$ is written as
\begin{equation}
\nabla R=-c\nabla t_{0}=\nabla _{1}R+(\partial R\mathbf{/}\partial
t_{0})\nabla t_{0},  \label{naber}
\end{equation}
which is Eq. (20-9) from$^{7}$ and the same is written in the unnumbered
equation preceding the numbered equation (63.7) in Ref. 6. Then the
equations which constitute the transformation of the differential operators
from the coordinates of the field point to those of the radiator are given
as
\begin{equation}
\partial \mathbf{/}\partial t=(R/s)\partial \mathbf{/}\partial t_{0},\quad
\nabla =\nabla _{1}+(\nabla t_{0})\partial \mathbf{/}\partial t_{0},
\label{difop}
\end{equation}
Eqs. (20-8) and (20-11) in Ref. 7. The differential operator $\nabla _{1}$
refers to the differentation with respect to, e.g., the first argument of
the function $f$ in Eq. (\ref{retco2}), that is, differentation at constant
retarded time $t_{0}$. Expressing this statement in another way we can say
that $\nabla _{1}$ denotes the differentation with respect to the arguments $%
x,y,z$ of the function $f,$ i.e., $R,$ (\ref{retco}) or (\ref{retco2}), (and
thus of the potentials $\varphi $ and $\mathbf{A}$ and the fields $\mathbf{E}
$ and $\mathbf{B}$ as well), on which $f$ (consequently $\varphi ,$ $\mathbf{%
A,}$ $\mathbf{E}$ and $\mathbf{B}$) \emph{explicitly depend} (not only
implicitly through $t_{0}$). This consideration reveals that the relations
(11) from Ref. 1 have to be correctly written as
\begin{eqnarray}
\partial \varphi /\partial x_{i} &=&(\partial \varphi /\partial
x_{i_{1}})+(\partial \varphi /\partial t_{0})(\partial t_{0}/\partial x_{i}),
\nonumber \\
\partial \mathbf{A}/\partial t &=&(\partial \mathbf{A}/\partial
t_{0})(\partial t_{0}/\partial t),  \label{derpot} \\
\partial A_{k}/\partial x_{i} &=&(\partial A_{k}/\partial
x_{i_{1}})+(\partial A_{k}/\partial t_{0})(\partial t_{0}/\partial x_{i}),
\nonumber
\end{eqnarray}
where $\partial /\partial x_{i_{1}}$ denotes, as above $\nabla _{1},$ the
differentation with respect to the argument $x_{i}$ (i.e., $x$ or $y$ or $z$%
) on which $\varphi $ or $A_{k}$ explicitly depend. The partial derivative,
e.g., $\partial /\partial x_{i}$ on the l.h.s. of (\ref{derpot}) can be
called - the \emph{complete} partial derivative - which takes into account
both the explicit and the implicit dependence of the functions $\varphi $ or
$A_{k}$ on the argument $x_{i}$. The partial derivative $\partial /\partial
x_{i_{1}}$ in the first and the third equation on the r.h.s. of (\ref{derpot}%
) takes into account only the \emph{explicit }dependence of the functions $%
\varphi $ or $A_{k}$ on the argument $x_{i}.$ For the sake of brevity we
shall, from now on, call such type of the partial derivative - the \emph{%
explicit} partial derivative.

When such procedure is applied to the calculation of $\mathbf{E}$ and $%
\mathbf{B}$ by Eq. (\ref{eandb}) from the LW potentials $\varphi $ and $%
\mathbf{A}$ (\ref{poten}) then one finds the exact expressions for the usual
LW fields (\ref{efiel}) and (\ref{bfiel}) without need for addition of any
new terms. We note that in Ref. 1 Eqs. (11) the partial derivatives which
refer to the explicit dependence of $\varphi $ and $\mathbf{A}$ on $x_{i}$,
i.e., $(\partial \varphi /\partial x_{i_{1}})$ and $(\partial A_{k}/\partial
x_{i_{1}}),$ were not taken into account$.$ As a consequence the authors of$%
^{1}$ had to add an additional term in their Eq. (21) to find the result (%
\ref{bfiel}). \bigskip

\noindent \textbf{3. The Retarded Electromagnetic Fields of a Charge Moving
with a Constant Acceleration Do Satisfy Maxwell Equations\medskip }

\noindent It is argued in Ref.1 that the electric and magnetic fields for a
charge $q$ moving with a constant acceleration along the $X$ axis, their
Eqs. (22)-(27), do not satisfy ME (\ref{maxw}). These fields are the
retarded fields (\ref{efiel}) and (\ref{bfiel}) but written in components,
and taking into account that the velocity and the acceleration of that
charge $q$ have only $x$-components, i.e., $\mathbf{V}(V,0,0)$ and $\mathbf{a%
}(a,0,0).$ We shall, for brevity, quote here only $E_{x}$ and $B_{y}$
components, Eqs. (22) and (26) from Ref. 1,
\begin{eqnarray}
E_{x}(x,y,z,t) &=&Kq\left\{ \left( (x-x_{0})-VR/c\right) \left(
1-V^{2}/c^{2}\right) /\left( R-V(x-x_{0})/c\right) ^{3}\right\} _{t_{0}}
\nonumber \\
&&+Kq\left\{ a\left( (x-x_{0})^{2}-R^{2}\right) /c^{2}\left(
R-V(x-x_{0})/c\right) ^{3}\right\} _{t_{0}},  \label{eiks}
\end{eqnarray}
\begin{eqnarray}
B_{y}(x,y,z,t) &=&-Kq\left\{ Vz\left( 1-V^{2}/c^{2}\right) /c^{2}\left(
R-V(x-x_{0})/c\right) ^{3}\right\} _{t_{0}}  \nonumber \\
&&-Kq\left\{ aRz/c^{3}\left( R-V(x-x_{0})/c\right) ^{3}\right\} _{t_{0}}.
\label{beips}
\end{eqnarray}
To prove the above mentioned assertion the authors of$^{1}$ supposed, as in
their Sec. 2., that $\mathbf{E}$ and $\mathbf{B}$ fields, their Eqs.
(22)-(27), are functions of $x,y,z,$ $t$ only through $t_{0},$ and
consequently used the differentiation rules as in their Eqs. (11). But from
our discussion in Sec. 2., see the relation for $R$ (\ref{retco2}), and from
the explicit expressions for $\mathbf{E}$ and $\mathbf{B}$ fields, (\ref
{eiks}) and (\ref{beips}) or Eqs. (22)-(27) from Ref.1, one unambiguously
concludes that $\mathbf{E}$ and $\mathbf{B}$ fields are also the explicit
functions of the field point $x,y,z.$ Thence the relations (28) and
(30)-(32) from Ref.1 have to be changed in accordance with our discussion
given in Sec. 2. Thus, e.g., the second relation in Eq. (28) in Ref. (1) has
to be changed and written as
\begin{equation}
\partial E\left\{ orB\right\} _{k}/\partial x_{i}=(\partial E\left\{
orB\right\} _{k}/\partial x_{i_{1}})+(\partial E\left\{ orB\right\}
_{k}/\partial t_{0})(\partial t_{0}/\partial x_{i}),  \label{eorbk}
\end{equation}
where the partial derivative on the l.h.s. of (\ref{eorbk}) is (in our
terminology) - the \emph{complete }partial derivative - which takes into
account both the explicit and the implicit dependence of the functions $%
E\left\{ orB\right\} _{k}$ on the argument $x_{i}$, while the partial
derivative in the first term on the r.h.s. of (\ref{eorbk}) is - the \emph{%
explicit} partial derivative, which takes into account only the explicit
dependence of the functions $E\left\{ orB\right\} _{k}$ on the argument $%
x_{i}.$ Of course the relations (30)-(32) from$^{1}$ have to be changed in
the same way as explained above.

Then one finds that the term $\nabla _{1}\times \mathbf{E}$ (due to the
explicit dependence of $\mathbf{E}$ on the arguments $x,y,z$) \emph{exactly
cancels }the terms on the r.h.s. of Eqs. (35) and (36) from Ref. 1. In such
a way we prove, as expected, that the Faraday law is exactly satisfied. It
can be easily shown that not only the Faraday law but all ME (\ref{maxw})
are exactly satisfied by the LW fields (\ref{efiel}) and (\ref{bfiel}), and
thus also by the fields from uniformly accelerated charge, (\ref{eiks}) and (%
\ref{beips}) or Eqs. (22)-(27) from Ref. 1, when all partial derivatives are
correctly treated.

At this place it is worth to discuss the results from Sec. 4 in Ref. 1,
where the similar mathematically incorrect step has been made. The authors of%
$^{1}$ argued that in the calculations of $\mathbf{E}$ and $\mathbf{B}$ from
(\ref{eandb}) and in ME (\ref{maxw}) as well all derivatives have to be
considered as \emph{total }ones. To explain the need for such change they
calculate $\partial t_{0}/\partial t$ and $\partial t_{0}/\partial x_{i}$,
Eqs. (14) in,$^{1}$ using two different expressions for $R$, our Eq. (\ref
{retco}), or Eqs. (37) and (38) from Ref. 1. Then to calculate the above
mentioned expressions they claim that the commonly used partial derivatives $%
\partial R/\partial t$ and $\partial R/\partial x_{i}$ have to be replaced
by the total derivative $dR/dt$ and $dR/dx_{i}$, see Eqs. (39) in Ref. 1.
Thus they also objected the use of the partial derivatives by Landau and
Lifshitz$^{6}$ (in the calculation of $\partial R/\partial t$ and $\partial
R/\partial x_{i}$) considering that Landau and Lifshitz$^{6}$ used the
symbol $\partial $: ''in order to emphasize that $R$ depends also on other
independent variables $x,y,z$,$...$ .'' The justification for such
assertions they find in the following statement: ''The point is that if a
given function is expressed by two different types of functional
dependences, then exclusively \emph{total} derivatives of these expressions
with respect to a \emph{given} variable can be equated (contrary to the
\emph{partial} ones).'' However this statement is not true. Namely (using
our terminology) it is true that one cannot equate - the \emph{explicit}
partial derivatives, but one can equate - the \emph{complete }partial
derivatives of the mentioned expressions. Moreover when one calculates,
e.g., $\partial R/\partial x,$ then it has to be written as
\begin{equation}
\partial R/\partial x=(\partial R/\partial x_{1})+(\partial R/\partial
t_{0})(\partial t_{0}/\partial x),  \label{comder}
\end{equation}
since according to (\ref{retco}) and (\ref{retco2}) $R$ depends explicitly
not only on $x$ but also on $y$ and $z$ (thence $\partial R/\partial x_{1}$%
). Further, we explain what is the most important for the use of the
operator $\partial /\partial x$ on the l.h.s. of (\ref{comder}) (also (\ref
{naber}), (\ref{difop}), (\ref{derpot}) and (\ref{eorbk})) and not the total
derivative operator $d/dx.$ It is the fact that $t_{0}$ in $R$ depends not
only on the $x$ variable but also on other independent variables $y$ and $z$%
. Of course $\partial R/\partial x$ on the l.h.s. of (\ref{comder}) is (in
our terminology) - the \emph{complete} partial derivative of $R$, while $%
\partial R/\partial x_{1}$ (the first term on the r.h.s. of (\ref{comder}))
is - the \emph{explicit }partial derivative of $R$. To see that this
explanation is correct and to understand the difference between the total
derivative and the partial derivative of the function (composite) of several
variables one can consult some mathematical book, e.g., Ref. 8 Eqs. (8) and
(9) in Sec. 153. Also one can see the nice explanation given in Ref. 7 Sec.
(20-1). This means that the objections from$^{1}$ to the Landau and Lifshitz$%
^{6}$ derivations and results are unfounded. Hence in Eqs. (39) from Ref. 1
the \emph{complete }partial derivatives $\partial R/\partial t$ and $%
\partial R/\partial x_{i}$ have to be written and not the \emph{total}
derivatives $dR/dt$ and $dR/dx_{i}$. Obviously the \emph{partial}
derivatives have to be retained in all other expressions and they must not
be replaced by the \emph{total} derivatives. The proposition from Ref. 1
that partial derivatives have to be replaced by the total derivatives has
been also extensively used for numerous conclusions in Refs. 2-4, see
particularly Eqs. (43)-(46) and the discussion in Ref. 2. We see that such
changes of ME\ are unnecessary and, in fact, incorrect. \bigskip

\noindent \textbf{4. Conclusion\medskip }

\noindent The consideration presented in this paper unambiguously reveals
that Maxwell equations (\ref{maxw}) are correctly written in the usual way
with the partial derivatives and not with the total derivatives, as argued
in Ref. 1. Furthermore, as proved in our Sec. 3., the retarded
Li\'{e}nard-Wiechert fields for an uniformly accelerated charge, (\ref{eiks}%
) and (\ref{beips}) or Eqs. (22)-(27) from Ref. 1, exactly satisfy the usual
form of ME (\ref{maxw}). Hence it is not true that: ''there is a \emph{%
simultaneous} and \emph{independent }coexistence of \emph{instantaneous} and
\emph{retarded interactions} which cannot be reduced to each other.'' The
retarded solution is a full-value solution, i.e., it is a complete solution
of ME, which is sufficient to describe the whole electromagnetic phenomenon.
In addition, it has to be mentioned (see also Ref. 5) that it is possible to
write the complete solution of ME in the present-time formulation, i.e., as
an action-at-a-distance formulation, and such a solution is completely
equivalent to the retarded (and advanced) solution. This is explicitly shown
in the general case in Ref. 9 by means of Lagrange series expansion. We
shall present the potentials and fields in the present-time formulation in a
\emph{closed form} for the case of an uniformly accelerated charge, but it
will be reported elsewhere. \bigskip

\noindent \textbf{References\medskip }

\noindent 1. A.E. Chubykalo and S.J. Vlaev, \textit{Int. J. Mod. Phys.}
\textbf{14}, 3789 (1999).

\noindent 2. A.E. Chubykalo and R. Smirnov-Rueda, \textit{Mod. Phys. Lett.}
\textbf{A12}, 1 (1997).

\noindent 3. A.E. Chubykalo, \textit{Mod. Phys. Lett.} \textbf{A13}, 2139
(1998).

\noindent 4. A.E. Chubykalo and R. Smirnov-Rueda, \textit{Phys. Rev.}
\textbf{E53}, 5373 (1996);

\textbf{E55}, 3793(E) (1997); \textbf{E57}, 3683(1998).

\noindent 5. T. Ivezi\'{c} and Lj. \v {S}kovrlj, \textit{Phys. Rev.} \textbf{%
E57}, 3680 (1998).

\noindent 6. L.D. Landau and E.M. Lifshitz, \textit{The Classical Theory of
Field }(Pergamon,

Oxford, 1975).

\noindent 7. W.K.H. Panofsky and M. Phillips, \textit{Classical Electricity
and Magnetism}

(Addison-Wesley, Reading, MA, 1962).

\noindent 8. V.I. Smirnov, Kurs Vis\v {s}ei Matematiki, Tom I (Nauka,
Moscow, 1967),

(English translation: A Course of Higher Mathematics, Vol I (Pergamon,

Oxford, 1964)).

\noindent 9. R.A. Villecco, \textit{Phys. Rev.} \textbf{E48}, 4008 (1993).

\end{document}